\documentclass[preprint,12pt]{elsarticle}

\usepackage{amssymb}
\usepackage{amsthm}

\usepackage{multirow}
 \usepackage{lineno}
\usepackage{url}
\usepackage{xcolor}
\usepackage{bm}
\usepackage{footnote}
\makesavenoteenv{table}
\usepackage{nicematrix}
\usepackage{threeparttable,array}
\makeatletter
\g@addto@macro{\UrlBreaks}{\UrlOrds}
\makeatother

\journal{Advances in Quantum Chemistry}

\begin{document}

\begin{frontmatter}



\title{Ionization potentials and electron affinity of oganesson}


\author[1]{Yangyang Guo}
\author[2]{Luk\'{a}\v{s} F. Pa\v{s}teka}
\author[3]{Ephraim Eliav}
\author[1]{Anastasia Borschevsky\corref{cor1}}
\address[1]{Van Swinderen Institute for Particle Physics and Gravity,
University of Groningen, Nijenborgh 4, 9747 Groningen, The Netherlands\\}
\address[2]{Department of Physical and Theoretical Chemistry \& Laboratory for Advanced Materials, Faculty of Natural Sciences, 
Comenius University, Ilkovi\v{c}ova 6, 84215 Bratislava, Slovakia\\}
\address[3]{School of Chemistry, Tel Aviv University, 6997801 Tel Aviv, Israel\\}
\cortext[cor1]{a.borschevsky@rug.nl}
\begin{abstract}
We present high accuracy relativistic coupled cluster calculations of the first and second ionization potentials and the electron affinity of the heaviest element in the Periodic Table, Og. The results were extrapolated to the basis set limit and augmented with the higher order excitations (up to perturbative quadruples), the Breit contribution, and the QED self energy and vacuum polarisation corrections. We have performed an extensive investigation of the effect of the various computational parameters on the calculated properties, which allowed us to assign realistic uncertainties on our predictions. Similar study on the lighter homologue of Og, Rn, yields excellent agreement with experiment for the first ionization potential and a reliable prediction for the second ionization potential.


\end{abstract}
 

\end{frontmatter}


\section{Introduction}
\label{}

Oganesson (Og), element 118, is the heaviest element in the Periodic Table. It was first synthesised at the Joint Institute for Nuclear Research (JINR) in Dubna \cite{PhysRevC.74.044602}, and in 2016, it received its name along with elements 113 (Nihonium), 115 (Moscovium), and 117 (Tennessine) \cite{KarBarShe16}. As the element that  completes the Periodic Table and is the gateway to elements with higher atomic numbers that have not been yet discovered, knowledge of the atomic properties of oganesson could lead to a better fundamental understanding of the effect of relativity on the trends in the electronic structure of the heaviest elements \cite{GiuMatNaz19}. However, to date, only a small number of Og atoms were produced, and, with a lifetime of around a single ms for the only isotope known so far ($^{294}$Og) \cite{BreUtyRyk18}, any experimental spectroscopic or chemical investigation of this element is currently out of our reach. Accurate and reliable theoretical investigations are thus presently the only route for obtaining information about this intriguing atom. A large number of recent high quality publications can be found addressing atomic, chemical, and bulk properties of oganesson on various levels of theory. 

Oganesson is assigned to Group 18 of the periodic table and is found under Rn, thus formally belonging to the rare gases. However, the huge impact of relativistic effects on its electronic structure (in particular, the large spin-orbit splitting of its $7p$ shell and the stabilisation of the vacant $8s$ orbital \cite{V.Pershina}) is expected to lead to properties that are uncharacteristic for the rare gases. In particular, a number of recent works predict Og to be a solid at room temperature \cite{JerSmiMew19,SmiMewJer20} and to exhibit semiconductor behaviour \cite{MewJerSmi19}.

An unusual atomic feature of Og is its electron affinity, which is predicted to be positive \cite{PhysRevLett.77.5350,GoiLabEli03,LacDzuFla18}, in contrast to its lighter homologues. Furthermore, the electron localization function of Og shows a uniform Fermi-gas-like behaviour in its valence region, somewhat smearing out its shell structure \cite{JerSchSch18}. The same work predicts a rather high polarizability of 57.98~a.u. for Og, which could lead to an increase in the Van der Waals interactions of this element and perhaps contribute to its curious bulk properties.

The examples above demonstrate that theoretical investigations provide us with a powerful framework that allows us to probe the otherwise inaccessible atomic, chemical, and even solid-state properties of this rare and short-lived element and provides us with an insight into its behaviour. When dealing with heavy systems, such studies should employ reliable computational methods that are based on relativistic approaches and that incorporate electron correlation effects on a high level. 

The aim of this work is to provide predictions of the basic atomic properties of Og obtained on the present highest possible level of theory. We thus use the relativistic coupled cluster approach with single, double, and perturbative triple excitations (DC-CCSD(T)) to calculate the first and the second ionization potentials and the electron affinity of Og. Furthermore, we correct our results for the higher order effects, such as the Breit and the QED contributions and the excitations beyond perturbative triples, following the scheme developed in Ref. \cite{PasEliBor17}. The extensive computational study that we perform is used to set uncertainties on our predictions, as we have recently done for the calculated electron affinity of At \cite{LeiKarGuo20}. In order to evaluate the accuracy of our predictions, and the reliability of the uncertainty estimates, we also perform calculations of the first ionization potential of the lighter homologue of Og, radon, where accurate experimental value is available \cite{NIST_ASD}. 
Setting an uncertainty on our predictions for Og will facilitate the use of the present results in further theoretical studies of this element and in possible future experiments. We extend this study to the second ionization potential of Rn, where the available experiment has a very large uncertainty \cite{Finkelnburg1955}.  

We conclude by comparing our results to the latest theoretical values obtained with high accuracy methods, which include relativistic Fock space coupled cluster (FSCC) approach \cite{PhysRevLett.77.5350,GoiLabEli03, HanDolHan12, JerSchSch18}, effective core potentials combined with CCSD(T) (ECP-CCSD(T)) \cite{Nas05}, DC-CCSD(T) \cite{PerBorEli08}, and relativistic configuration interaction approach combined with many-body perturbation theory (CI+PT) \cite{LacDzuFla18}.

\clearpage
\section{Methods and computational details}

All the calculations were carried out in a relativistic framework, based on the Dirac-Coulomb (DC) Hamiltonian (in atomic units): 

\begin{eqnarray}
H_{DC}= \displaystyle\sum\limits_{i}h_{D}(i)+\displaystyle\sum\limits_{i<j}(1/r_{ij}),
\label{eqHdcb}
\end{eqnarray}
where $h_D$ is the relativistic one-electron Dirac Hamiltonian,
\begin{eqnarray}
h_{D}(i)=c\bm{\alpha}_{i}\cdot \bm{p}_{i}+c^{2}\beta _{i}+V^n(i).
\label{eqHd}
\end{eqnarray}

Here  $\alpha$ and $\beta$  are the four dimensional Dirac matrices, $V^n(i)$ is the nuclear attraction operator, and c is the speed of light. The nuclear Coulomb potential $V^n(i)$ takes into account the finite size of the nucleus modelled by a Gaussian charge distribution \cite{VisDya97}. The no-virtual-pair approximation \cite{Suc80} is based on the restricted kinetic balance approach \cite{DyaFae90}.

The starting point of our investigations were the mean-field Dirac--Hartree--Fock (DHF) calculations. For the neutral atoms we used the closed-shell DHF, while for the open-shell systems the average of configuration (AOC) type calculations were performed \cite{Roothaan1960,McWeeny1974,Thyssen2004}. For the singly and doubly charged Rn and Og, we thus represented the open-shell system with 4 and 5 valence electrons, respectively, that were evenly distributed over 6 valence p spinors; in case of the Og anion, the valence electron was allowed to occupy the 8s orbital. The electron correlation was taken into account using the relativistic coupled-cluster approach with single, double and perturbative triple excitations, DC-CCSD(T). All the electrons were correlated and the virtual orbitals were cut-off at 300 a.u. A calculation with a higher cutoff of 2000 a.u. was carried out to to check that the size of correlation space is sufficient, and to evaluate the uncertainty due to this truncation. All the relativistic CCSD(T) calculations were carried out using the DIRAC17 program package \cite{DIRAC17}.

We have used the relativistic preconstructed correlation-consistent Gaussian-type all-electron basis sets of Dyall \cite{Dya06,Dyall2012} and explored the performance of different quality sets (double-, triple- and quadruple-zeta quality, designated v2z, v3z and v4z, respectively). In addition, also the cv$N$z and ae$N$z basis sets were used, which include additional high angular momentum (high $l$) functions with high exponents that are needed to correlate the core-valence region and inner core electrons, respectively. Results were  extrapolated to the complete basis set (CBS) limit, using the usual Dunning--Feller $e^{-\alpha N}$ scheme \cite{Dunning1989,Feller1992} for the DHF values and the popular $N^{-3}$ CBS scheme of Helgaker \textit{et al.}\cite{Helgaker1997} for the correlation contribution, following our previous studies \cite{PasEliBor17,LeiKarGuo20}.
Alternative CBS extrapolation schemes were tested and the results were used in our CBS error estimation. For the mean-field (DHF) extrapolation, we tested the scheme of Karton and Martin \cite{Karton2005} and for the extrapolation of the correlation contributions we tested the $(N+\frac{1}{2})^{-4}$ scheme of Martin \cite{Martin1996} and the recent more involved scheme of Lesiuk and Jeziorski \cite{Lesiuk2019} based on the rigorous analysis of correlation in He-like systems. 

\begin{figure}[t]
  \centering
  \includegraphics[width=0.7\linewidth]{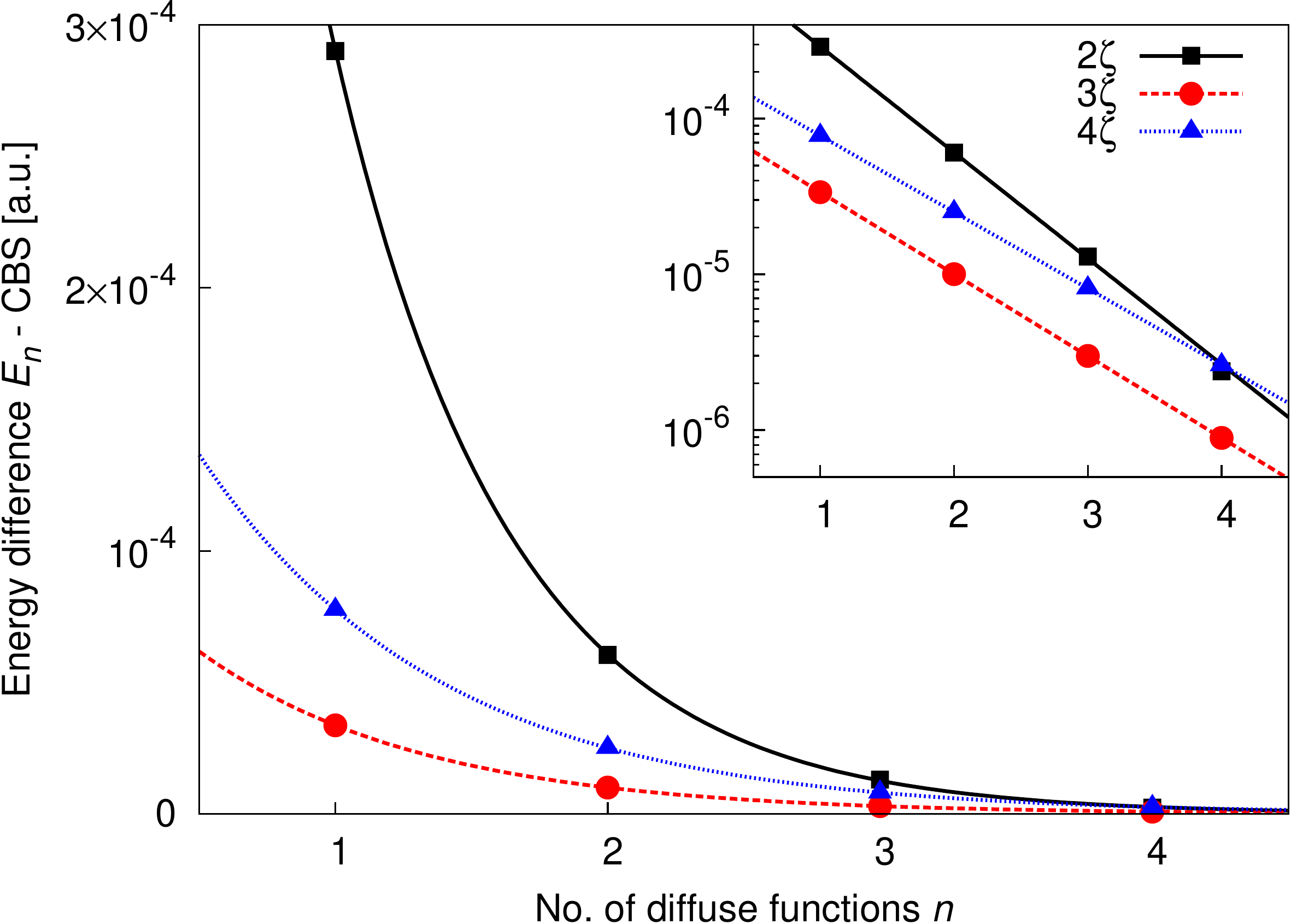}
  \caption{Exponential convergence of the total CCSD(T) atomic energy $E_n$ of Og towards the CBS limit with the increasing level of basis set augmentation $n$ for ($n$-aug)-cv$N$z basis sets. Semi-log inset showcases perfect linearity.}
  \label{fig:diffuse}
\end{figure}

In order to obtain quantitatively correct results for the electron affinity of a loosely bound anion, high quality description of the region distant from the nucleus (which will contain the added electron) is necessary. We have thus augmented the basis sets with 3 diffuse functions for each symmetry block. The first layer of these functions was optimised in the original basis sets \cite{Dya06,Dyall2012}, and designated as (1-aug)-cv$N$z/ae$N$z, respectively; further augmentation layers were generated automatically, in an even-tempered fashion, and designated (2-aug)-, (3-aug)-, etc. 
Furthermore, we observed near-perfect exponential asymptotics with the increasing number of even-tempered diffuse functions (Figure \ref{fig:diffuse}). This allowed us to extrapolate the total energies to infinite augmentation limit (denoted (($\infty$-aug)-cv$N$z) using a simple exponential function analogous to the Dunning--Feller scheme \cite{Dunning1989,Feller1992}.
A similar systematic augmentation expansion was used earlier in the context of the EA of methane \cite{RamrezSols2014}.

In order to improve the accuracy of our predictions and go beyond the relativistic CCSD(T) approximation we have included higher order effects in our calculations. In terms of electron correlation, this means including excitations beyond perturbative triples. 

Higher-order correlation is dominated by the valence contributions \cite{PasEliBor17}. Full T and perturbative (Q) contributions were thus calculated in a smaller correlation space containing the valence electrons (Rn: 5d, 6s, 6p; Og: 6d, 7s, 7p) with a virtual orbital cutoff of 14 a.u. and 16 a.u. for Og and Rn, respectively, corresponding to equally sized correlation spaces. These calculations were performed with the program package MRCC \cite{MRCC,KalSur01,BomStaKal05,KalGau05,KalGau08} linked to DIRAC15 \cite{DIRAC15} using Dyall's augmented valence av\textit{X}z basis sets \cite{Dya06}. The DIRAC15 version was used due to its compatibility with the MRCC code. The $\Delta$T contributions were extrapolated to the CBS limit as above and an additional core-valence correction to $\Delta$T was calculated at the (1-aug)-cv2z level.
$\Delta$(Q) contributions, showing almost no basis set dependence, were calculated at the (1-aug)-v3z level. 

We next turn to improving the treatment of relativity in our calculations. The one-electron part of the DC Hamiltonian is relativistic, while the Coulomb operator can be considered as a non-relativistic description of the 2-electron interaction. Since the non-instantaneous interactions between the electrons are limited by the speed of light in the relativistic framework, we add the frequency-independent Breit correction to the two-electron part of $H_{DC}$. This interaction is given by

\begin{eqnarray}
B_{ij}=-\frac{1}{2r_{ij}}[\bm{\alpha }_{i}\cdot \bm{\alpha }_{j}+\frac{(%
\bm{\alpha }_{i}\cdot \bm{r}_{ij})(\bm{\alpha }_{j}\cdot \bm{%
r}_{ij})}{r_{ij}^{2}}],
\label{eqBij}
\end{eqnarray}
in the Coulomb gauge. 

To further improve precision we include also the QED corrections in the form of the model Lamb shift operator (MLSO) of Shabaev and co-workers \cite{ShaTupYer15}. This model Hamiltonian uses the Uehling potential and an approximate Wichmann--Kroll term for the vacuum polarisation (VP) potential \cite{BLOMQVIST197295} and local and nonlocal operators for the self-energy (SE), the cross terms (SEVP), and the higher-order QED terms \cite{ShaTupYer15}.
Both the Breit term and the Lamb shift were calculated with the Tel Aviv atomic computational package \cite{TRAFS-3C} using the Fock-space coupled cluster method (DCB-FSCC) and the extended universal basis sets \cite{MalSilIsh93}, consisting of $37s$, $31p$, $26d$, $21f$, $16g$, $11h$, and $6i$ functions.

The calculated higher order excitation contributions and the Breit and QED corrections were added to the DC-CCSD(T) results to obtain the final recommended values of the ionization potentials and the electron affinity. The comprehensive computational investigation that we performed allows us to set uncertainties on our predictions, following the procedure presented in Ref. \cite{LeiKarGuo20} and outlined below.\

 \section{Results and discussion}

\subsection{Basis set effects}

Table \ref{tab:basis} presents the investigation of the effect of gradually enlarging the basis set on the calculated first and second IPs of Rn and Og and the EA of Og; the presented values were obtained at the CCSD(T) level of theory. 

The effect of using the all-electron basis set ae4z, compared with the core-valence basis set cv4z is between 3 and 7 meV only. This difference decreases slightly for the (1-aug)-ae$4$z vs (1-aug)-cv4z comparison. We thus continue with the core-valence basis sets for the remainder of this work. 

The calculated ionization potentials of both elements show very little effect of addition of diffuse functions beyond (1-aug)-cv4z level (that is, beyond the first augmentation layer). 
The electron affinity of Og, on the other hand, only becomes positive for the (2-aug)-cv4z basis set, and almost doubles in value for the (3-aug)-cv4z basis. This shows the importance of the diffuse functions, which describe the area distant from the nucleus, for high quality description of binding of an electron to a neutral closed shell atom. 

Extrapolation to the complete basis set limit (with respect to the cardinality number $N$) has a moderate effect of about a 100 meV for the ionization potentials and 9 meV for the EA of Og. Extrapolation to the infinitely augmented basis leads to a further 4 meV increase in the EA of Og, while the IPs remain unchanged.

\begin{table}[t]
  \centering
    \caption{Calculated IPs and EA of Rn and Og using varying quality basis sets. The calculations were carried out on relativistic CCSD(T) level of theory. }
    \begin{tabular}{@{\extracolsep{4pt}}l c c c c c@{}}
    \hline\hline
    \multirow{2}{*}{Basis set}&\multicolumn{2}{c}{Rn}& \multicolumn{3}{c}{Og}\\
    \cline{2-3}\cline{4-6}
    &IP$_1$&IP$_2$&IP$_1$&IP$_2$&EA\\
   \hline
cv3z & 10.465&18.683 &8.627	&15.922	&--2.994  \\
 cv4z &10.629&18.857& 8.755	&16.079	&--2.092 \\ 
  ae4z &10.624& 18.854 &8.756&	16.084&	--2.085 \\ 
  (1-aug)-cv4z& 10.659 &18.877 & 8.791	&16.092	&--0.223 \\ 
  (1-aug)-ae4z&10.655&18.875&8.791&16.097&	--0.221
\\
  (2-aug)-cv4z &10.659 &18.877 & 8.791&	16.092&	\phantom{--}0.040  \\
  (3-aug)-cv4z&10.659 &18.877&8.791&	16.092&	\phantom{--}0.069 \\ 
  (3-aug)-CBS-cv$N$z&10.772 &19.001 & 8.882 & 16.197 & \phantom{--}0.078 \\ 
  $\infty$-CBS-cv$N$z &10.772 &19.001 & 8.882 & 16.197 & \phantom{--}0.082 \\
      \hline\hline
  \end{tabular}
  \label{tab:basis}
\end{table}

\begin{figure}[t]
  \centering
  \includegraphics[width=\linewidth]{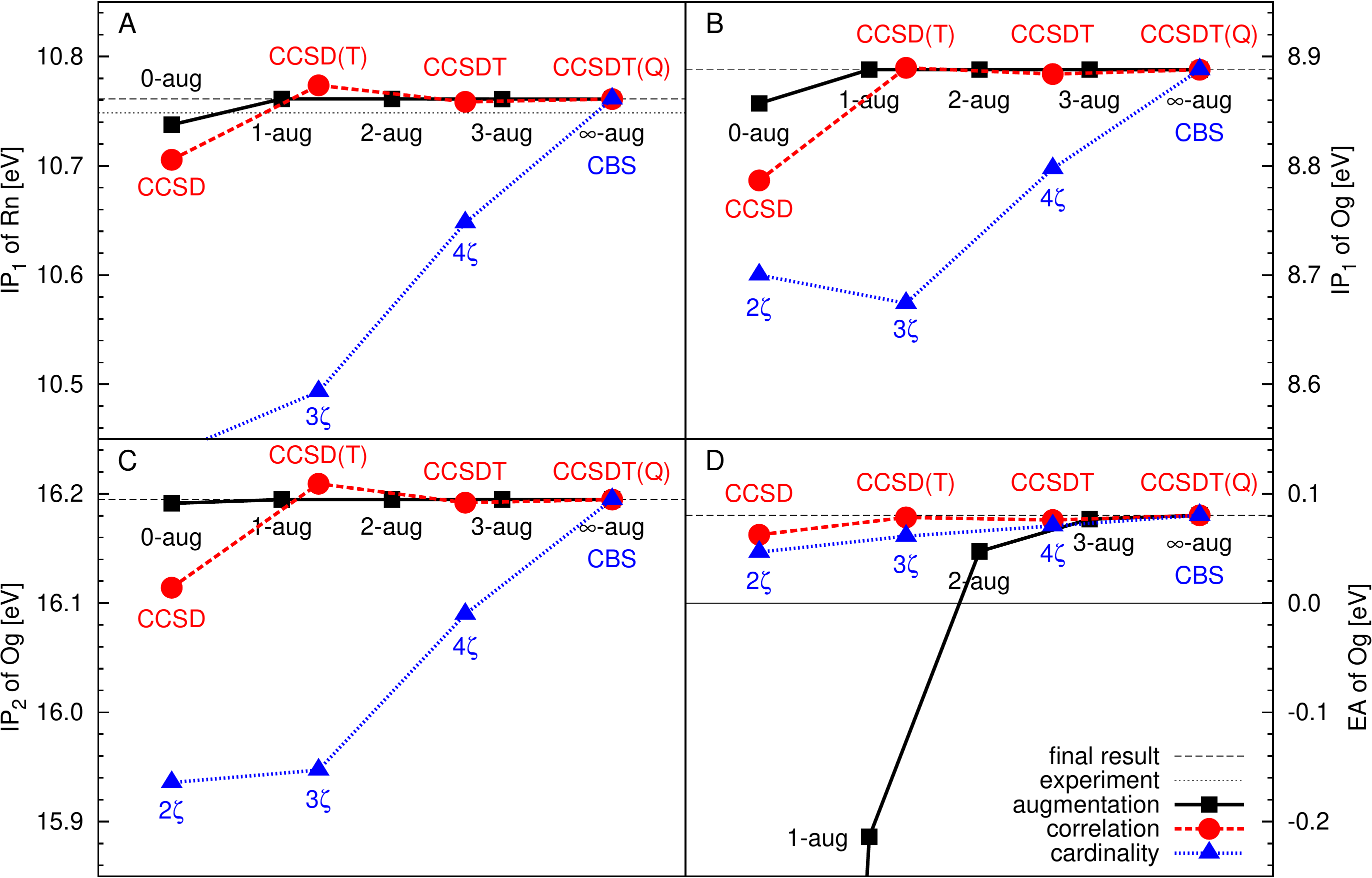}
  \caption{Effect of the basis set cardinality, the level of augmentation, and the treatment of electron correlation on the calculated IP$_1$(Rn) (A), IP$_1$(Og) (B), IP$_2$(Og) (C), EA(Og) (D). The black dotted lines represent the final result extrapolated to the complete basis set and infinite augmentation limit, corrected for all higher order contributions. The dash-dot line in (A) represents the experimental IP$_1$(Rn)}
  \label{fig:RnOgIPEA}
\end{figure}

Plots in Figure \ref{fig:RnOgIPEA} visually summarize the trends in the basis set and correlation effects on the calculated IPs and EA (on a comparable 0.4 eV scale). The latter effects are discussed in the following subsection. For the former, we observe that the effect of basis set quality is dramatically different in terms of cardinality and augmentation when calculating IPs compared to the EA of Og. A single layer of diffuse functions is sufficient for full convergence of all the IP results. In turn, IPs depend significantly more on the basis set cardinality. For the EA of Og, the situation is reversed - even at double-zeta level one obtains a quite satisfactory result provided the basis set is sufficiently diffuse. Since the excess electron is only loosely bound in Og$^-$, extensive basis set augmentation is necessary. Coupled with systematic extrapolation to infinitely augmented basis set, we improve the confidence of the EA result not being an artifact of electron confinement due to suboptimally diffuse basis set. 

\subsection{Electron correlation and other corrections}

The contributions of the various higher order corrections are listed in Table \ref{tab:high} and the electron correlation trends are shown in Figure \ref{fig:RnOgIPEA}.

The perturbative triple contributions are of the order of 70$-$100 meV for the IPs of both elements, and about 20 meV for the EA of Og. In the latter case, these are particularly important, as the total EA is quite small. 
Going from the perturbative to the full triple excitations decreases the IPs and the EA, while the quadruple excitations increase the results by several meV; the higher order correlation contributions thus partially cancel out. Overall, the effect of correlation beyond CCSD(T) is notably smaller than is often encountered for the elements outside the rare gas group. This is also easily seen in the convergence trends in Figure \ref{fig:RnOgIPEA}.

The Breit contribution is quite small in all cases and affects the second IPs the most. The QED contributions are of similar magnitude.

All the higher order correction terms are added on top of the CBS extrapolated CCSD(T) values from Table \ref{tab:basis} to obtain the final results presented in section \textit{3.4}. For these elements and properties, the total contribution of the higher order effects do not exceed $\sim$10 meV. 

\begin{table}[t]
  \centering
    \caption{Contributions of the higher order corrections (meV). $\Delta$(T) shown for comparison.}
    \begin{tabular}{@{\extracolsep{4pt}}l r r r r r @{}}
    \hline\hline
    \multirow{2}{*}{Sources}&\multicolumn{2}{c}{Rn}& \multicolumn{3}{c}{Og}\\
    \cline{2-3}\cline{4-6}
    & \multicolumn{1}{c}{IP$_1$} & \multicolumn{1}{c}{IP$_2$} & \multicolumn{1}{c}{IP$_1$} & \multicolumn{1}{c}{IP$_2$} & \multicolumn{1}{c}{EA}\\
   \hline
   $\Delta$(T)&68.0 &72.1 & 103.1 & 95.2& 15.9\\
    $\Delta$T & --15.1 & --8.0 & --5.7 & --17.3 &--2.8  \\ 
  $\Delta$Q& 2.7 & 0.8 & 4.0 & 2.8 & 4.7 \\ 
Breit &--1.7& --9.0 & 1.4 & --1.8 & --0.3 \\
 QED &3.1 &5.3 & 6.5 & 14.0 & --3.0 \\ 
      \hline\hline
  \end{tabular}
  \label{tab:high}
\end{table}

\subsection{Uncertainties}

Uncertainty estimations based on accurate computational investigations allow us to set error bars on our predicted values for the second IP of Rn and the IPs and EA of Og. Such error bars are important for the planning of future measurements of these properties. 
There are different sources of error arising from the various approximations employed in the calculations, which can be broadly divided into uncertainties due to the finite basis set size, the incomplete treatment of correlation, and the missing relativistic effects. We treat these sources of error separately and assume them to be largely independent.  

The present calculations are based on the core-valence (1-aug)-cv$N$z basis set family, due to the excessive computational demands of using the larger all-electron (1-aug)-ae$N$z basis sets. 
Nevertheless, all electrons were correlated in our calculations. 
We evaluate the error due to the lack of the explicit extra core-correlating functions by taking the difference between the results obtained with the (1-aug)-ae4z and the (1-aug)-cv4z basis sets, which amounts to several meVs (Table \ref{tab:errors}, Core correlation functions).

\begin{table*}[t]
 \centering
    \caption{Summary of the main sources of uncertainty in the calculated IPs and EA of Rn and Og (meV).}
    \begin{tabular}{@{\extracolsep{4pt}}l l r r r r r@{}}
    \hline\hline
    \multirow{2}{*}{Category}&\multirow{2}{*}{Error source} &\multicolumn{2}{c}{Rn}& \multicolumn{3}{c}{Og}\\
    \cline{3-4}\cline{5-7}
    & & \multicolumn{1}{c}{IP$_1$} & \multicolumn{1}{c}{IP$_2$} & \multicolumn{1}{c}{IP$_1$} & \multicolumn{1}{c}{IP$_2$} & \multicolumn{1}{c}{EA}\\
    \hline
 \multirow[t]{3}{*}{Basis set}&core corr. functions & --4.4& --1.6&0.3  &4.5  & 1.9 \\
 &CBS &57.1&	65.2&	43.2&	48.9&3.2  \\ 
 & augmentation & 0.0&0.0 & 0.0& 0.0 &1.8  \\[1.5ex]
\multirow[t]{3}{*}{Correlation} & virtual cutoff &--0.1 &0.0 & --0.1 & --0.1 & 0.2 \\
  &higher excitations &2.0  &0.9& 3.1 & 4.3 & 2.4  \\ [1.5ex]

 Relativity& QED &3.1 &5.3 & 6.5 & 14.0 & --3.0\\[1.5ex]
  Total&&57.4& 65.4 &43.8 &51.3 & 5.7  \\ 
      \hline\hline
  \end{tabular}
    \label{tab:errors}
\end{table*}

The scheme of Helgaker \textit{et al.}\cite{Helgaker1997} (H) stood the test of time and is arguably the most popular CBS extrapolation approach found in the literature. We have  successfully used it in our earlier studies \cite{PasEliBor17,LeiKarGuo20} and hence we also use it here. However, to evaluate the uncertainty of the CBS extrapolation, we have also employed two alternative schemes introduced in the previous section: the scheme of Martin \cite{Martin1996} (M), and of Lesiuk and Jeziorski \cite{Lesiuk2019} (LJ). Table \ref{tab:cbs} contains the extrapolated results using the different schemes. 
When comparing the different schemes we observe that the LJ results are consistently higher than H values while the M results are lower by a similar amount. This almost uniform behavior further justifies us to use the central H extrapolation scheme for the final values obtained in this work. However, the spread in these values allows us to determine their standard deviation ($\sigma$) and hence also the typical 95\% confidence interval (1.96$\sigma$) as our CBS uncertainty estimate.

The uncertainty due to the incomplete description of the region distant from the nucleus is estimated by taking a conservative half of the difference between the (3-aug)-CBS-acv$N$z and the extrapolated ($\infty$-aug)-CBS-acv$N$z results. This uncertainty is negligible for the ionization potentials but more significant for the electron affinity of Og.

\begin{table}[t]
  \centering
    \caption{Spread in the CCSD(T)/(3-aug)-cv$N$z results obtained using different schemes for the extrapolation to the complete basis set limit and the corresponding uncertainty estimation. See text for further details.}
    \begin{tabular}{@{\extracolsep{4pt}}l r r r r r @{}}
    \hline\hline
    \multirow{2}{*}{Scheme}&\multicolumn{2}{c}{Rn}& \multicolumn{3}{c}{Og}\\
    \cline{2-3}\cline{4-6}
    & \multicolumn{1}{c}{IP$_1$} & \multicolumn{1}{c}{IP$_2$} & \multicolumn{1}{c}{IP$_1$} & \multicolumn{1}{c}{IP$_2$} & \multicolumn{1}{c}{EA}\\
   \hline
LJ-CBS &10.801&	19.032&	8.905&	16.224&	0.082\\
H-CBS &10.772&	19.001&	8.882&	16.197&	0.079\\
M-CBS &10.743&	18.966&	8.861&	16.174	&0.079\\
95\% c.i. & $\pm$0.057 & $\pm$0.065 &	$\pm$0.043 &	$\pm$0.049 & $\pm$0.003\\
      \hline\hline
  \end{tabular}
  \label{tab:cbs}
\end{table}

Next, we turn to the errors stemming from the treatment of the electron correlation, namely the effect of cutting off the virtual correlation space and of neglecting excitations beyond (Q).  
To estimate the effect of the virtual cutoff, we take the difference between the results obtained with the cutoff of 300 a.u., which we use for the final values presented in this work, and those with a cutoff of 2000 a.u. (using the cv4z basis set). In all cases this contribution to the uncertainty is found to be negligible. 

Similarly to the base value, to estimate the uncertainty for the higher excitation contributions, we combine the $\Delta$T CBS extrapolation error (determined as above), together with the virtual cutoff error (difference with respect to the 40 a.u. cutoff), and the neglect of the contributions beyond the perturbative quadruples. The latter was estimated as a half of the $\Delta$(Q) contribution itself based on the typical decreasing hierarchy of CC contributions exemplified also by the comparison between the $\Delta$(T) and $\Delta$T contributions shown in Table \ref{tab:high}. This choice was further justified by $\Delta$Q contributions calculated at the (1-aug)-v2z level being all below 1meV.

We assume the higher order QED corrections will not be larger than the vacuum polarization and the self energy contributions.

The total uncertainty is obtained by combining all the above terms and assuming them to be independent. As we are treating uncertainties due to the higher order effects, such an assumption is reasonably justified. As can be seen in Table \ref{tab:errors}, the dominant source of error for all calculated properties is the extrapolation of the basis set to the CBS limit. This could be significantly reduced by introducing a pentuple-$\zeta$ basis set into the extrapolation. Unfortunately, Dyall's basis set family currently only contains basis sets up to the quadruple-$\zeta$ level.

\subsection{Final values}

Table \ref{tab:final} contains the final values of the first and second ionization potentials of Rn and Og, and the electron affinity of Og, including the associated uncertainties, and compared to experiment for the first ionization potential of Rn and with earlier theoretical predictions for the rest of the values. We focus on theoretical works that either investigate Og or both Og and Rn, rather than papers presenting information on Rn only.  Our calculated IP$_1$ of Rn is in excellent agreement with the experimental value \cite{doi:10.1021/ja059868l}, with a difference of only 0.17\% and well within our estimated uncertainties. This lends confidence to both the predicted values and the associated uncertainties for the IP$_2$ of Rn and all the calculated properties of Og, where no experiment is available. 

To the best of our knowledge, the second IP of Rn has not been yet studied using high-accuracy computational methods and the present work provides a reliable prediction for this property. It should be noted that this prediction is somewhat at odds with the 1955 experimental value \cite{Finkelnburg1955}, which could perhaps motivate new measurements of this property.

 The previous predictions of the IP of Og are between 8.842 eV with the FSCC method and 8.866 eV obtained with CI+PT (except for the earlier ECP+CCSD(T) calculations \cite{Nas05} based on a limited quality basis set), and the present value is 20 meV above this interval, probably due to the use of extended basis sets and extrapolation to the complete basis set limit.
 Our calculated electron affinity of Og is somewhat higher than the earlier FSCC predictions but lower than the CI+PT result; all the previous values lie outside the present uncertainty. The advantage of the present work over the earlier investigations is mainly in the systematic description of the diffuse basis functions (as compared to the FSCC calculations) and the inclusion of the excitation beyond singles and doubles (as compared to all earlier works).

\begin{figure}[ht!]
  \centering
  \includegraphics[width=0.8\linewidth]{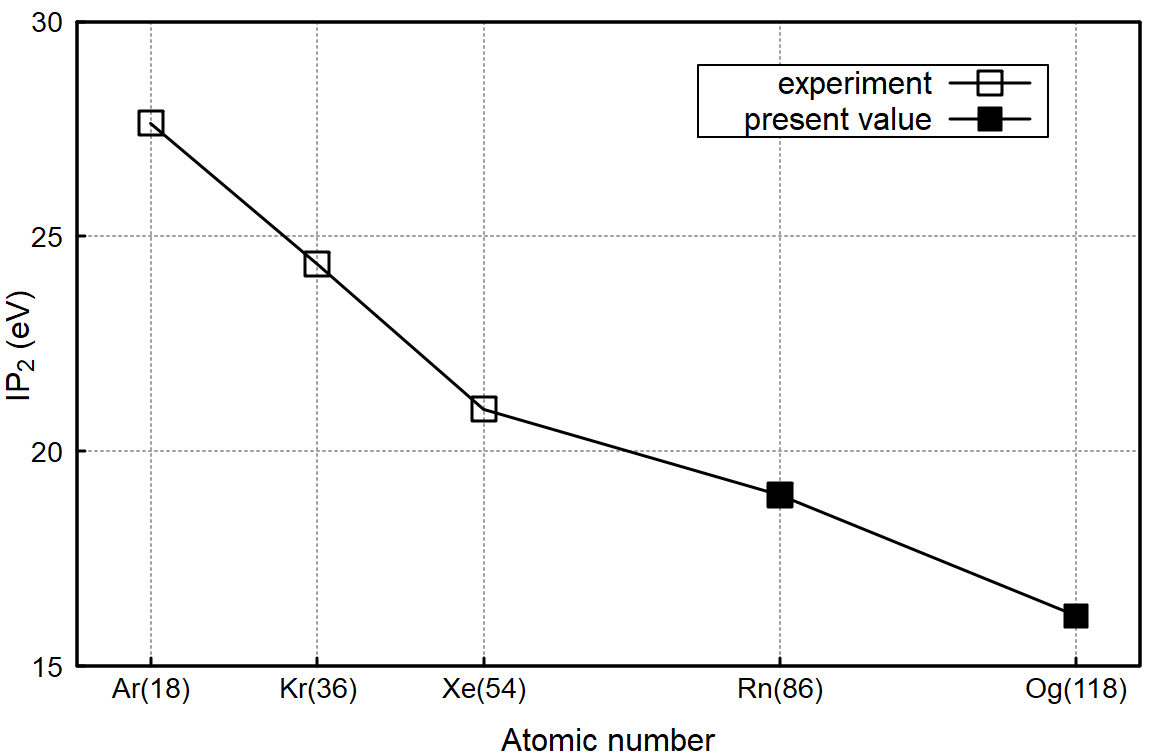}
  \caption{Second ionization potentials IP$_{2}$ of the rare gases (eV); the values of Ar to Xe are taken from experiment and those for Rn and Og from the present calculations.}
  \label{fig:IP12}
\end{figure}

The trend in the first ionization potentials of the rare gases, including Og was presented in Ref. \cite{PerBorEli08} (Figure 1(b)). Here, we show a similar plot for the second IPs of the rare gases, as shown in Fig. \ref{fig:IP12}; the values for Ar to Xe are taken from experiment \cite{doi:10.1063/1.555896, Hansen_1987, Ar-IP2} while for Rn and Og we present the current predictions. The decrease in the second IP when going from Rn to Og is slightly steeper than between Xe and Rn, due to the destabilisation of the valence  $7p_{3/2}$ orbital in the heavier ion. 

\begin{table*}[ht!]
  \centering
    \caption{IP$_1$, IP$_2$ and EA of Rn and Og (in eV), compared with experimental and earlier theoretical results.}
    \resizebox{\linewidth}{!}{
    \begin{tabular}{@{\extracolsep{4pt}}l l l l l l c@{}}
    \hline\hline
     \multirow{2}{*}{Method}&\multicolumn{2}{c}{Rn}& \multicolumn{3}{c}{Og}&\multirow{2}{*}{Ref.}\\
    \cline{2-3}\cline{4-6}
    & \multicolumn{1}{c}{IP$_1$} & \multicolumn{1}{c}{IP$_2$} & \multicolumn{1}{c}{IP$_1$} & \multicolumn{1}{c}{IP$_2$} & \multicolumn{1}{c}{EA\phantom{XX}}\\
   \hline

CBS-CCSDT(Q) & 10.761(57) & 18.990(65)&	8.888(44)&16.195(51)&0.080(6) & Present\\
\phantom{X}+Breit+QED & &&&	& & \\
 FS-CCSD+Breit &  &  & & & 0.056(10)& \cite{PhysRevLett.77.5350} \\ 
FS-CCSD+Breit &  & & & &0.064(2) &  \cite{GoiLabEli03}\\ 
\phantom{X}+QED & & &&	&& \\
 CI+PT& 10.876&&8.866&  & 0.096 & \cite{LacDzuFla18}\\ 
 
FSCC& & &8.842&  &  &\cite{JerSchSch18}\\
ECP-CCSD(T)&10.482&& 8.642&&&\cite{Nas05}\\
FSCC&&& 8.864&&&\cite{HanDolHan12}\\
DC-CCSD&10.799&&8.863&&&\cite{PerBorEli08}\\
Exp.&10.7485&  21.4(19)&&&&\cite{doi:10.1021/ja059868l,NIST_ASD,Finkelnburg1955}\\
      \hline\hline
  \end{tabular}}
  \label{tab:final}
\end{table*}
\section{Conclusions}

We have carried out high accuracy calculations of the first and second ionization potentials and the electron affinity of Og. The relativistic CCSD(T) approach was used in the calculations and we corrected the results for higher excitations (up to perturbative quadruples) and for the Breit and lowest-order QED contributions. Extensive basis set investigation was performed, and the results were extrapolated to the complete basis set level both in terms of cardinality and in terms of augmentation with diffuse functions. In the later case, we tested and applied a new extrapolation scheme. The computational study that we performed allowed us to assign realistic error bars on our results. An accompanying study on the first ionization potential of Rn yielded a result that is in excellent agreement with the experimental value (and well within the estimated uncertainties), supporting our predictions for its  heavier homologue, Og. 

Accurate and reliable predictions of the basic atomic properties of this rare and short lived element, accompanied by realistic uncertainties, are important in supporting future experimental and theoretical research at the edge of the Periodic Table. 

\section{Acknowledgement}
We would like to thank the Center for Information Technology of the University of Groningen for providing access to the Peregrine high performance
computing cluster and for their technical support.
LFP acknowledges the support from the Slovak Research and Development Agency (APVV-20-0098, APVV-20-0127) and the Scientific Grant Agency of the Slovak Republic (1/0777/19).

\clearpage
  \bibliographystyle{naturemag}
  \bibliography{bib}





\end{document}